# Low-density functionalized amorphous carbon nanofoam as binder-free Thin-film Supercapacitor electrode

Subrata Ghosh[1,2*], Massimiliano Righi[1], Andrea Macrelli[1], Francesco Goto[3], Marco Agozzino[1], Gianlorenzo Bussetti[3], Valeria Russo[1], Andrea Li Bassi[1], Carlo S. Casari[1*]

[1] *Micro and Nanostructured Materials Laboratory – NanoLab, Department of Energy, Politecnico di Milano, via Ponzio 34/3, Milano, 20133, Italy*

[2] *Warsaw University of Technology, Faculty of Mechatronics, św. A. Boboli 8, Warsaw 02-525, Poland*

[3] *Solid Liquid Interface Nano-Microscopy and Spectroscopy (SoLINano-Σ) lab, Department of Physics, Politecnico di Milano, Piazza Leonardo da Vinci 32, 20133 Milano, Italy*

Nanoporous carbon materials containing small domains of *sp*$^2$-carbon with highly disordered structures are promising for supercapacitor applications. Herein, we synthesize amorphous carbon nanofoam with 98% volumetric void fraction and low mass density of around 30 mg/cm$^3$ by pulsed laser deposition at room temperature. With the unavoidable oxygen functional groups on the nanoporous surface, carbon nanofoam and nitrogen-functionalized carbon nanofoams are directly grown on the desired substrate under different background gases (Ar, $N_2$, $N_2$-$H_2$), and employed as supercapacitor electrodes. Among the background gases used in synthesis, the use of nitrogen yields nanofoam with higher thickness and more N-content with higher graphitic-N. From the test of amorphous carbon nanofoam supercapacitor device, nitrogenated amorphous carbon electrode shows a higher areal capacitance of 4.1 mF/cm$^2$ at 20 mV/s in aqueous electrolyte, a better capacitance retention at higher current, and excellent cycle stability (98%) over 10000 charge-discharge cycles are achieved compared to not-functionalized counterpart prepared under Ar background gas (2.7 mF/cm$^2$ and cycle stability of 88%).

*Keywords: Nanocarbons, amorphous, nanofoam, pulsed laser deposition, plasma plume, supercapacitor, Energy Storage*

Corresponding author email: subrata.ghosh@polimi.it (S.G.) and carlo.casari@polimi.it (C.S.C.)

ORCID ID

0000-0002-5189-7853 (S. G.); 0000-0002-5307-5124 (A.M.); 0000-0001-8556-8014 (G.B.); 0000-0001-9543-0422 (V.R.); 0000-0002-1265-4971 (A. L. B.); 0000-0001-9144-6822 (C.S.C.)

## 1. Introduction

Amorphous carbon (*a*-C) has been broadly investigated fundamentally due to its typical characteristics including the presence of all *sp*, $sp^2$ and $sp^3$ hybridization, low-range order, huge amounts of defects, and plenty of active sites for electrochemical reactions, etc.[1][2][3] Depending upon the $sp^2/sp^3$ contribution and hydrogen content, *a*-C is mainly categorized into graphite-like, diamond-like, and polymer-like.[4] Each category shows unique optical, mechanical, and electrical properties, and is extensively used for energy storage, electronics and other applications.[1][5] A recent study anticipates freestanding monolayer amorphous carbon for magnetic recording devices and flexible electronic applications.[6] Among the carbon nanostructures, *a*-C nanofoams (NFs) have received significant attention due to their porous structure with high surface area, huge void fraction,[7] low mass density,[8][9] hydrophilicity,[7] unconventional magnetism[10][11], excellent thermal properties.[12] Moreover, porous carbon NF is explored as a suitable active material for Na-ion storage,[13] supercapacitors,[14] catalysis,[15] solar steam generation,[16] and Faradaic desalination of brackish water.[17] It is also used as a target in ion acceleration schemes by interaction with ultra-intense laser pulses[18][19], as mechanical platform to decorate nanoparticles such as Pt/Pd for electrochemical detection of cancer cells or Au and Ag to enhance the charge storage performance for supercapacitor applications.[20]

Carbon NF were first synthesized by pulsed laser deposition (PLD).[21][22] Although several synthesis techniques such as hydrothermal methods,[8] sol-gel methods,[23] pyrolysis[24], laser processing of a graphene oxide followed by transfer of the film[25] etc. are employed to obtain this unique nanostructure [12][26], they have their own limitations. For example, an ultralightweight carbon NF prepared by hydrothermal method needs 48 hours of synthesis at a temperature more than 165 °C,[27] while high temperature is one of the essential parameters to obtain NFs using pyrolysis method[24]. PLD has the advantage of growing the material at room temperature within relatively low deposition time and on any substrate [7]. To obtain porous structures like carbon NF in PLD, deposition is carried out at high deposition pressure and substrates are placed outside the shock wave front of plasma plume. Moreover, it has been shown that the mass density of carbon NF can be controlled by tuning the deposition gas type and pressure.[9] Till now, in most cases, ablation of a graphite target is performed in inert gases (Ar, He etc.).[9][3] while the role of other gases such as nitrogen and hydrogen on the deposition of *a*-C NF has not been explored in detail.

Heteroatoms (e.g. nitrogen, sulphur, phosphorous) doping into the carbon matrix is one of the common adoptable strategies to enhance the charge-storage performance of electrode material.[28] It has been reported that N-doped carbon NF derived from amino acid chelate complex exhibited an excellent charge-storage performance and is used for solar energy harvesting applications.[14] Also, nitrogen incorporation into the *a*-C matrix or nitrogenated *a*-C (*a*-C:N) showed a significant effect on the capacitance, work function and semiconductor properties.[29][28] Femtosecond pulsed laser deposited *a*-C:N film showed excellent electron transfer kinetics, and is anticipated as a promising material for electrochemical detection of electroactive pollutants.[30] In our previous report,[7] we showed that nitrogenated graphite-like *a*-C NF, grown by PLD at room temperature, could be a potential candidate for electrochemical energy storage application. However, the role of hydrogen- and nitrogen-content in the structure of *a*-C NF and its impact on the potential application such as supercapacitor was not addressed.

To elucidate this role, the present report focuses on the deposition of pristine and functionalized *a*-C NF directly on the substrates (silicon and carbon paper current collector) by pulsed laser deposition at room temperature. The pristine *a*-C NF is deposited under Ar background gas, whereas functionalized *a*-C NFs are prepared under $N_2$ and $N_2$-$H_2$, named *a*-C:N NF and *a*-C:N:H NF, respectively, keeping other synthesis parameters constant. The growth rate, thickness and mass density and structural properties of those NFs are thoroughly investigated and correlated with the plasma plume chemistry under different background gases. Finally, as-grown NF is employed as a binder-free and conductive additive-free electrode for aqueous electrochemical energy storage applications.

## 2. Experimental methods

### *2.1. Synthesis methodology*

*a*-C NF were prepared by PLD using graphite (1 inch diameter) as the target with a purity of 99.99% (purchased from Testbourne B. V.). The graphite target was placed on a rotating/translating holder, while the Si (100) and carbon paper substrates were placed on the rotating substrate holder at a distance of 38 mm from the target. The *ns*-PLD setup exploits the second harmonic ($\lambda$ = 532 nm) of a Q-switched Nd:YAG laser (pulse duration of 5–7 ns), with a repetition rate of 10 Hz and a fluence of 2.7 J/cm$^2$. Prior to the deposition, the chamber was evacuated down to the base pressure of $2 \times 10^{-3}$ Pa using first a rotary pump followed by a turbomolecular pump. Thereafter, desired background gases, Ar, $N_2$ and 95% $N_2$-5% $H_2$, were injected into the deposition unit. The deposition pressure was maintained at 500 Pa for each 15 min deposition. The deposition unit was vented, and the as-deposited NFs were taken out to examine the morphology, structure, and electrochemical properties as supercapacitor electrodes.

### *2.2. Microscopy and Spectroscopy.*

A field-emission scanning electron microscope (FE-SEM, ZEISS SUPRA 40, Jena, Germany) was employed to probe the morphology of carbon NF. Energy Dispersive X-ray (EDX) spectroscopy was used at the acceleration voltage of 5 kV to evaluate the local chemical composition, using a Peltier-cooled silicon drift detector (Oxford Instruments) and the Aztec software for quantification. A specific MATLAB code (named EDDIE software) was used to calculate the mass density (*r*) of NF with the input of area under Si, carbon and oxygen contents, an average thickness of the NF, and the details of each element such as atomic number and mass number.[31] The volumetric void fraction of carbon NFs was estimated using the relation of $\%\,porosity = \left(1 - \frac{\rho_{a-C\,NF}}{\rho_{graphite}}\right) \times 100$, where density of graphite is ($\rho_{graphite}$) of 2.2 g/cm$^3$.

Surface elemental compositions and bonding environment of NFs were examined by X-ray photoelectron spectroscopy (XPS) using a non-monochromatized X-ray source (Mg anode, photon energy 1253.6 eV), maintained at a power of 200 W. The kinetic energy of the photoemitted electrons was measured using a hemispherical analyzer with a 150 mm mean radius, PHOIBOS150 from SPECS GmbH. The spectra were acquired with a pass energy of 20 eV, with an energy resolution of 0.9 eV (full width at half maxima, FWHM). The pressure in the measurement chamber during the XPS measurements was about $1\times10^{-10}$ Torr. The fitting details are provided in our previous report.[7] Briefly, peaks were fitted after Shirley background subtraction using CasaXPS software, and at.% of elemental compositions were extracted from peak area ratios after correction by Scofield relative sensitivity factors (C = 1.0, N = 1.77, O = 2.85).[32] For C1s, the asymmetric $sp^2$-C peak is fitted with Gaussian-Lorentzian lineshape (GL (30)) with asymmetric factor (T200) and other symmetric carbon peaks with GL(30) by setting the range of FWHM to 1.2–2 eV. The FWHM of oxygenated carbon peaks, deconvoluted O1*s* peaks, and deconvoluted N1*s* peaks are set to 1.8-2.2 eV. The $sp^3$-C peak is shifted by 0.7-1 eV from $sp^2$-C and hydroxyl/ether, carbonyl and carboxylic groups are shifted approximately 1.5, 3, and 4.5 eV higher, respectively.

A Renishaw *InVia* Raman spectrometer was used to collect the spectra of carbon NFs. Each Raman spectrum was recorded using the 514.5 nm excitation radiation from an $Ar^+$ laser source with a power of 0.4 mW, an 1800 line/mm grating spectrometer and a 50× objective lens, with 20 accumulations for 10 s each.

### *2.3. Electrochemical measurements.*

The electrochemical performances of the NFs were investigated in a 2-electrode configuration using Swagelok Cell (SKU: ANR-B01, Singapore), basic 6M KOH (ACS reagent, Sigma-Aldrich, ≥85%) used as the aqueous electrolyte. The cell was assembled by sandwiching modified separator-soaked-electrolytes between two symmetric carbon NFs grown directly on carbon paper (Figure S2 of supplementary Material). Prior to the test, electrodes and modified separator were dipped into the electrolyte solution for 1h, and the cyclic voltammetry was conducted within the electrochemical stable voltage range of 0-0.8 V at 100 mV/s for 400 cycles. Cyclic voltammograms at different scan rates ranging from 20 to 1000 mV/s and charge-discharge at different currents from 125 to 500 μA were recorded using a Palmsens4 electrochemical workstation (PALMSENS, The Netherland). The areal capacitance was calculated using the equation: $C_{areal} = \frac{\int IdV}{A.v.V}$, where $I$ is the current, $v$ is the scan rate, $A$ is the geometric area of the electrode and $V$ is the voltage of the device. The areal capacitance of the device from charge-discharge profile was estimated using the relation of $C_{areal} \frac{I_d.t_d}{(A.\ V)}$, where $I_d$ and $t_d$ are the discharge current and the discharge time, respectively. Single electrode capacitance = 4 x device capacitance. The volumetric capacitance of electrode materials is estimated by dividing the areal capacitance by the total thickness of both electrodes. Energy density ($E$) and power density ($P$) of devices were estimated via the relations of $E = 0.5\ C_{areal}V^2$ and $P = {(E.v)}/{V}$, respectively.

## 3. Results and Discussions

### 3.1. *Morphological investigation:*

Typical top-view and cross-sectional scanning electron micrographs of NFs are displayed in Figure 1. Morphology-wise, there are no distinguishable changes that can be seen for the NFs grown in different background gases. The estimated mass densities ($\rho$) of the *a*-C, *a*-C:N and *a*-C:H:N NF are 32.6, 43.3 and 42.6 mg/cm$^3$, respectively (Table 1). This fact reflects that the NFs containing nitrogen are similar to each other but different from pristine ones (*a*-C). Secondly, the NF can be categorized as an ultralightweight material, whose density is much lower than the heavy carbons (mass density > 1 g/cm$^3$ for, e.g., diamond, graphite, and *a*-C) and other carbon nanostructures (with mass density in the range of 100-300 mg/cm$^3$ for, e.g., carbon nanotubes, nanoporous carbons, and carbon aerogels).[8] Figure 1 also confirms that NF is highly porous in nature. The estimated volumetric void fraction of all three NFs is around 98%, which is noticeably higher than the other carbon-based nanostructures such as carbon nanosheets prepared by plasma-enhanced chemical vapour deposition (71%).[33] Such amount of porosity offers higher electrolyte ions accessibility to the electrode surface leading to high charge-storage performance.[34] A noticeable difference between the NFs synthesized under different background gases is the thickness. The deposition rate for *a*-C, *a*-C:N, and *a*-C:N:H NF is found to be 0.85, 2.01 and 1.84 mm/min, respectively. The difference in growth rate of NF under the different background gases is attributed to the nature of background gas used during the deposition, as discussed below.

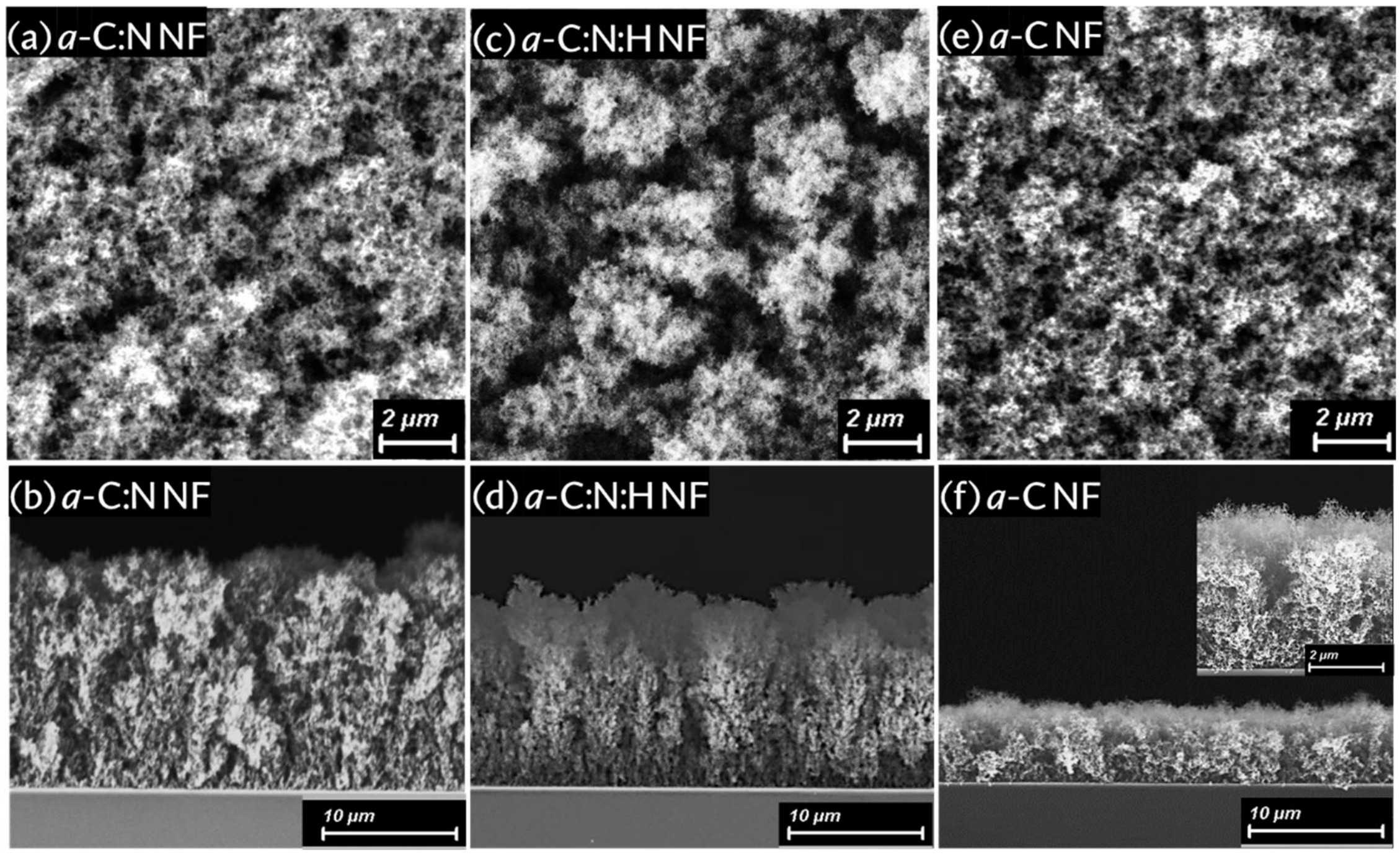

Figure 1: Scanning electron micrograph of carbon nanofoam. (a-b) *a*-C:N NF, (c-d) *a*-C:N:H NF, and (e-f) *a*-C NF.

Upon laser irradiation, as the plasma plume forms and expands, the ablated species move towards the substrate, and finally get deposited on its surface. During ablation in a high-pressure background gas, a shock wave front forms and propagates until the so-called stopping distance is achieved, resulting in plume spatial confinement, and enhanced visible glowing related to the number of collisional events occurring between ablated and gas species.[35] Thus, it is obvious that the plume propagation length, flux distribution, evolution of shock waves, plasma dynamics, kinetic energies, and/or arrival rates of the various constituents within the plume, and backward motion of ablated species are deeply affected by the background gas, chiefly in terms of chemical reactivity and atomic weight. Therefore, the type of background gas plays a crucial role in determining the structural quality, deposition rate, thickness, and mass density of deposited materials. In turn, one can visualize the shape evolution of the time-integrated plasma plume from Figure 2. The distance between the cloud center and the shock wave font is estimated to be 2.96, 3.22 and 3.37 cm for *a*-C, *a*-C:N, and *a*-C:N:H NF, respectively. Ar possesses a heavier molecular weight (atomic mass of 39.948 amu), yielding higher cloud density and hence a shorter propagation of the shock wave front compared to the lighter gas ($N_2$, atomic mass of 14.007 amu). At the same time, the backward motion of ablated species becomes more dominant when a heavier gas is used. To validate this assumption, we placed the substrates on the target-side and studied the impact of back-ablated species. It has been evidenced that the *a*-C NF deposited on target side under Ar (*a*-C NFT) has the maximum thickness of 29.5 mm (Figure S1). On the other hand, the maximum thickness for *a*-C:N NFT and *a*-C:N:H NFT is found to be 14.9 and 11.2 mm, respectively (Figure S1). Eventually, we have seen in our previous work that the deposition of the ablated species in the backside is the result of ballistic aggregation under highly reactive species, while the diffusive aggregation occurs for the film deposited on the frontside substrate.[7] Compared to a previous report where carbon NF was deposited at 300 Pa,[7] here the carbon NF was deposited at 500 Pa in order to obtain highly porous structure. Therefore, the investigation on the deposition formed by backward ablated species (Figure 2d-f) indicates that use of lighter mass is advantageous to lower the backward motion of ablated species and to synthesize the nanostructure on the frontside with higher yield. However, the lower deposition rate of *a*-C:N:H NF compared to *a*-C:N NF could be due to the etching of *sp*-C, $sp^2$-C and $sp^3$-C by atomic hydrogen at different rates along with the possibility of CH-like species formation inside the plasma plume.[36] Further detailed investigation on the plasma diagnostic is the subject of research.

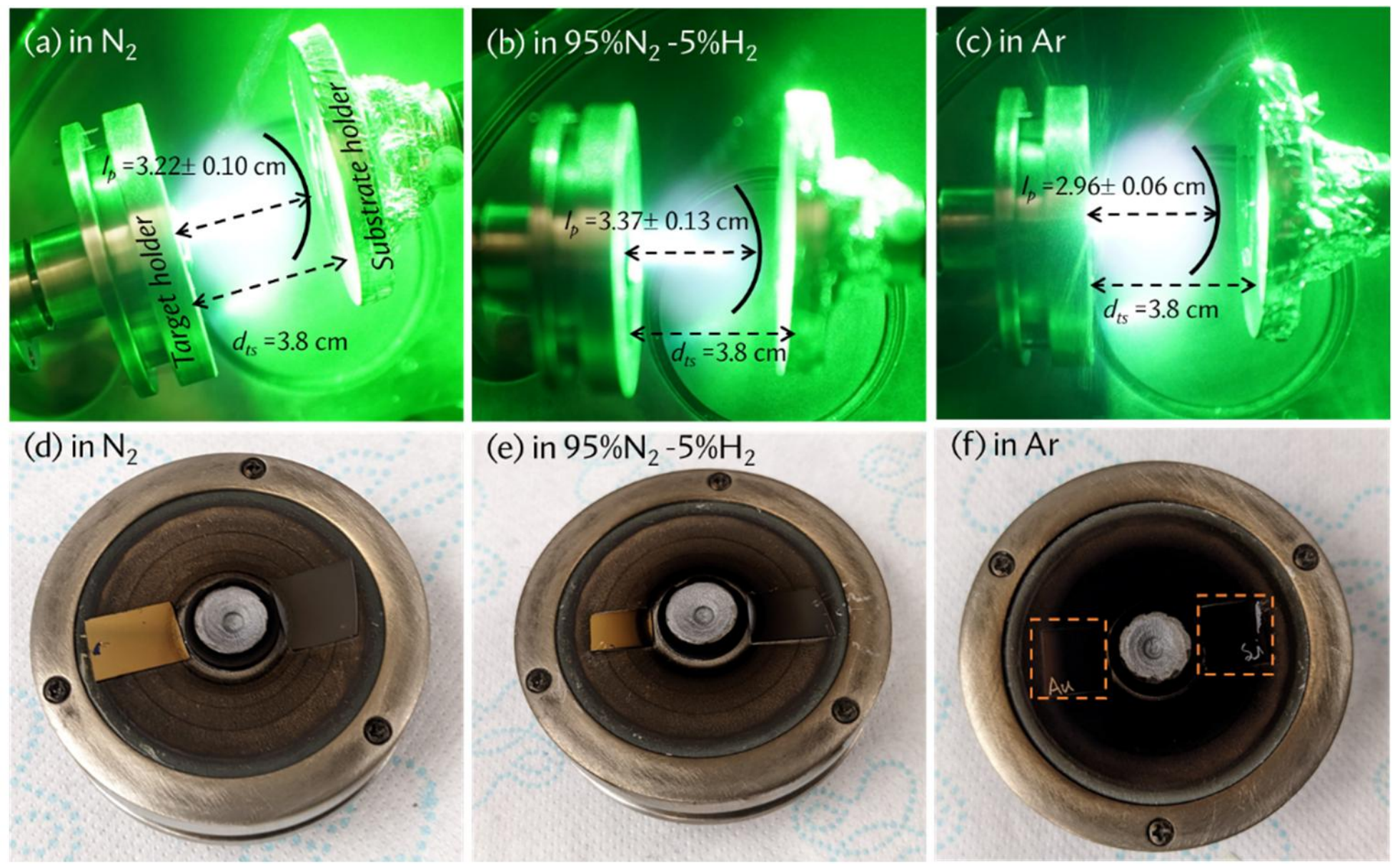

Figure 2: Digital photographs of plasma plume for the nanofoam grown in (a) $N_2$, (b) $N_2$-$H_2$ and (c) Ar. The ablation occurs on the small graphite target by green laser with a fluence of 2.7 J/cm$^2$ and the background pressure of 500 Pa in all cases. (see experimental section for details). Optical photograph of target after the growth of nanofoams in (d) $N_2$, (e) $N_2$-$H_2$ and (f) Ar.

### 3.2. *Structural investigation*

The Raman spectra of the three NFs are shown in Figure 3a. All the spectra present typical $sp^2$ features in the 1300-1600 cm$^{-1}$ range, namely D-peak and G-peak, and a broad second order Raman band in the region 2800-3000 cm$^{-1}$ where also CH bonds, when present, give contribution.[37] The three NFs, all prepared by ablating graphite in PLD under a background gas at room temperature, are amorphous as expected. Compared to *a*-C NF and *a*-C:N NF, a higher photoluminescence (PL) background is noticed for the *a*-C:N:H NF grown in $N_2$-$H_2$ background gas, suggesting the incorporation of hydrogen in the carbon matrix.[38] The minimal PL background of *a*-C and *a*-C:N NFs may be attributed to the adsorbed water upon exposure in ambient environments or poor vacuum during the deposition. No evident differences in the Raman spectrum of NFs grown under different background gas can be seen in Figure 3a. Anyway, a more quantitative analysis, based on fitting with deconvoluted peaks after background subtraction, reveals some structural differences among them (Figure 3b). The D-peak was fitted with a Lorentzian lineshape, while the G-peak with a Breit-Weigner-Fano lineshape, and the position and FWHM of the G-peak, and the intensity ratio D-to-G peaks ($I_D/I_G$) are reported in Table 1. Comparatively, *a*-C NF shows the lowest G-peak position with the largest FWHM (1575 cm$^{-1}$ and ~ 138 cm$^{-1}$, respectively), accompanied by $I_D/I_G$ estimated to be 0.51. The G-peak position for both *a*-C:N and *a*-C:N:H is graphitic-like (about 1583 cm$^{-1}$), with similar FWHM, lower than for *a*-C NF, suggesting a larger amount of $sp^2$ clustering probably induced by nitrogen [36]. On the other hand, the intensity ratio, $I_D/I_G$, is higher for *a*-C:N NF than *a*-C:N:H NF (0.71 and 0.54, respectively), suggesting a lower disorder for the $sp^2$ clusters in presence of hydrogen. This fact is attributed to the nitrogen-induced $sp^2$-clustering [39].

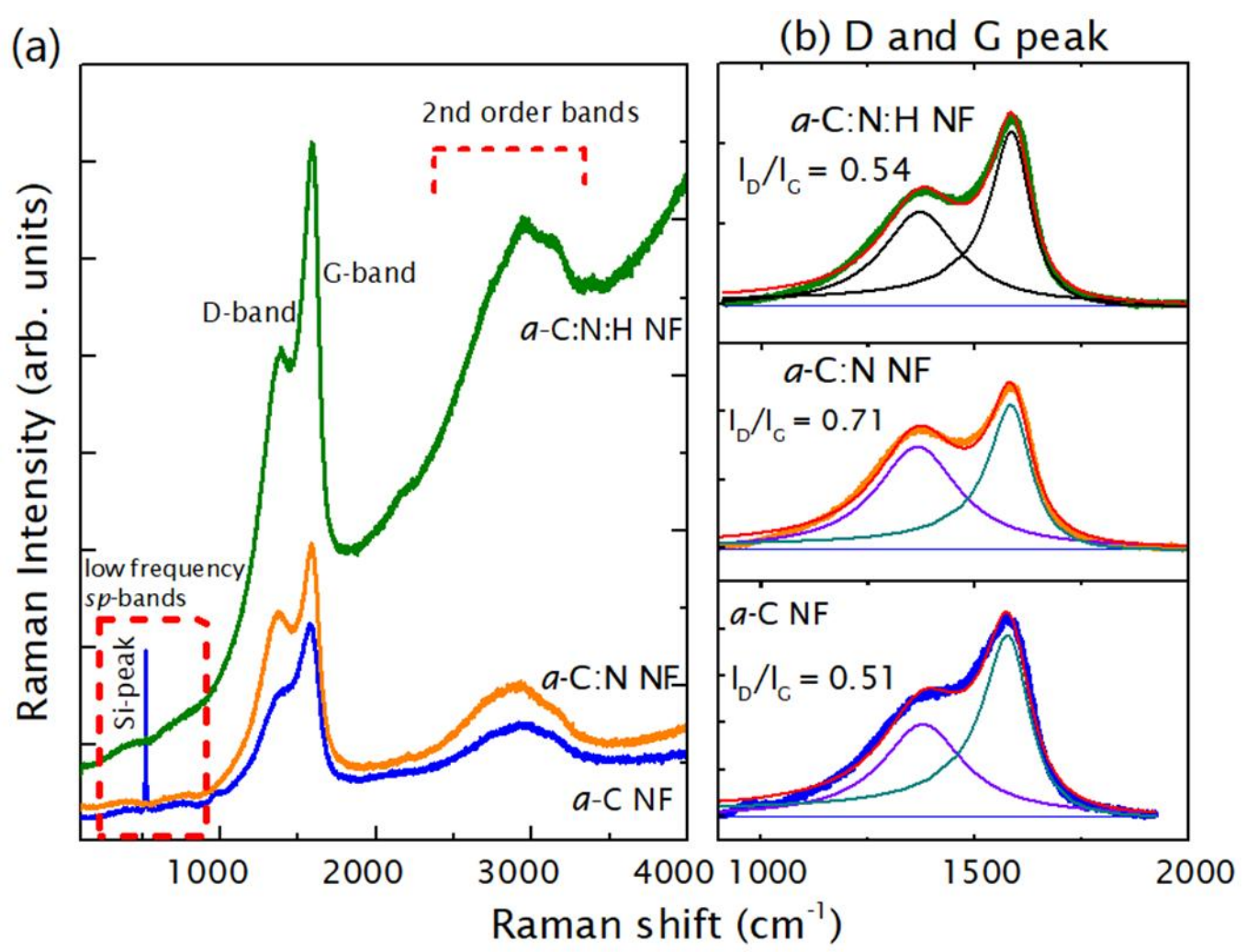

Figure 3 (color online): Visible Raman spectra of (a) *a*-C, *a* -C:N and *a* -C:N:H nanofoams grown on Si substrate. (b) Magnified 1st order Raman spectra of nanofoams with deconvoluted peaks.

Elemental compositional of NFs is investigated by both bulk EDX (Figure 4a, and Figure S3 of supplementary Material) and surface-sensitive XPS technique (Figure 4b-c). Among the NFs, a higher amount of carbon content is obtained for *a*-C:N NF compared to the other two NFs from both the techniques (Table 1). While EDX results indicate that *a*-C NF contains only carbon and oxygen (Figure S3 of supplementary Material), XPS showed the presence of 2.6 at% of nitrogen along with them (Figure 4c). The results obtained indicate that nitrogen and oxygen are physiosorbed on highly porous *a*-C NF since the deposition was carried out under pure Ar gas environment. On the other hand, both the EDX and XPS result for *a*-C:N NF reflect the relatively higher nitrogen incorporation in the *a*-C matrix than the pristine NF. Although the N-content is hardly quantified from the EDX spectra of *a*-C:N:H NF, a closer look reveals a very weak signature of nitrogen (Figure S3 of supplementary Material). The result obtained here about the role of hydrogen is also in good agreement with the deposition rate, average thickness, mass density and graphitic quality of NF as obtained from the Raman spectroscopic results.

Table 1: Physico-chemical and electrochemical properties of amorphous carbon nanofoams.

| Nanofoam (Background gas used in PLD) | Morphological result | | | Raman extracted parameters | | | XPS result | | | Charge-storage properties of device | | |
|---|---|---|---|---|---|---|---|---|---|---|---|---|
| | Avg. Thickness (μm) | Mass density (mg/cm³) | Vol. void fraction (%) | G-peak Position (cm⁻¹) | G-peak Full width at half maximum (cm⁻¹) | Intensity ratio of D-to-G peak | C/O/N (at.%) | $\frac{sp^2}{sp^3}$ | Graphitic-N (at.%) | Areal (vol.) capacitance in mF/cm² (F/cm³) at 20 mV/s | rate performance (%) at 1000 mV/s | cycle stability after 10000 cycles at 125 μA (%) |
| *a*-C:N ( $N_2$) | 10.1 | 43.3 | 98 | 1583 | 125 | 0.71 | 60.1/32.1/7.8 | 2.86 | 57.2 | 4.1 (2) | 22 | 98 |
| *a*-C:N:H(95%$N_2$-5%$H_2$) | 9.2 | 42.6 | 98 | 1584 | 118 | 0.54 | 57.5/39.4/3.1 | 1.88 | 32.7 | 1.2 (0.7) | 18 | 88 |
| *a*-C (Ar) | 4.3 | 32.6 | 98.5 | 1575 | 138 | 0.51 | 55.2/42.2/2.6 | 2.56 | 35.7 | 2.7 (3.1) | 36 | 73 |

To gain deeper inside on the bonding between carbon, nitrogen and oxygen, high-resolution XPS spectra are deconvoluted. A high resolution C1*s* spectra of *a*-C:N NF with deconvoluted peaks is shown in Figure 4(d), which consists of *sp*-C (at around 282.6 eV) [40] $sp^2$-C (at around 284.6 eV), $sp^3$-C (at around 285.7 eV), C-O/C-N (at around 286.4 eV), C=O/C=N (at around 288.4 eV), and -COO (at around 290.4 eV).

The high-resolution spectra of each element for *a*-C and *a*-C:N:H NF are provided in Figure S4 of the supporting file. The $sp^2$-C/$sp^3$-C ratio is reported in Table 1 for the three NFs, and it is found to be higher for *a*-C:N NF (2.86) compared to that of *a*-C NF (2.56) and *a*-C:N:H NF (1.88). A higher $sp^3$-carbon content is likely for *a*-C:N:H NF as it is deposited in presence of hydrogen, which can form C-H bonds with $sp^3$ hybridization. Incorporation of hydrogen in the NF is also confirmed by the high PL background in the Raman spectrum. The high resolution N1*s* spectra of NFs is also inspected to probe the different bonding of nitrogen, such as pyrrolic, pyridinic, graphitic and $NO_x$ and satellite peak (Figure 4e and S4). It has been seen that *a*-C:N NF contains higher graphitic nitrogen compared to the other two NFs (Figure 4f). This can further confirm the graphitic-like $sp^2$ clustering when only nitrogen is present during the deposition. All the NFs also have COOH, COH, C=O and N-C=O functional groups, and variable relative content in the NFs (Figure 4g-h and Figure S4).

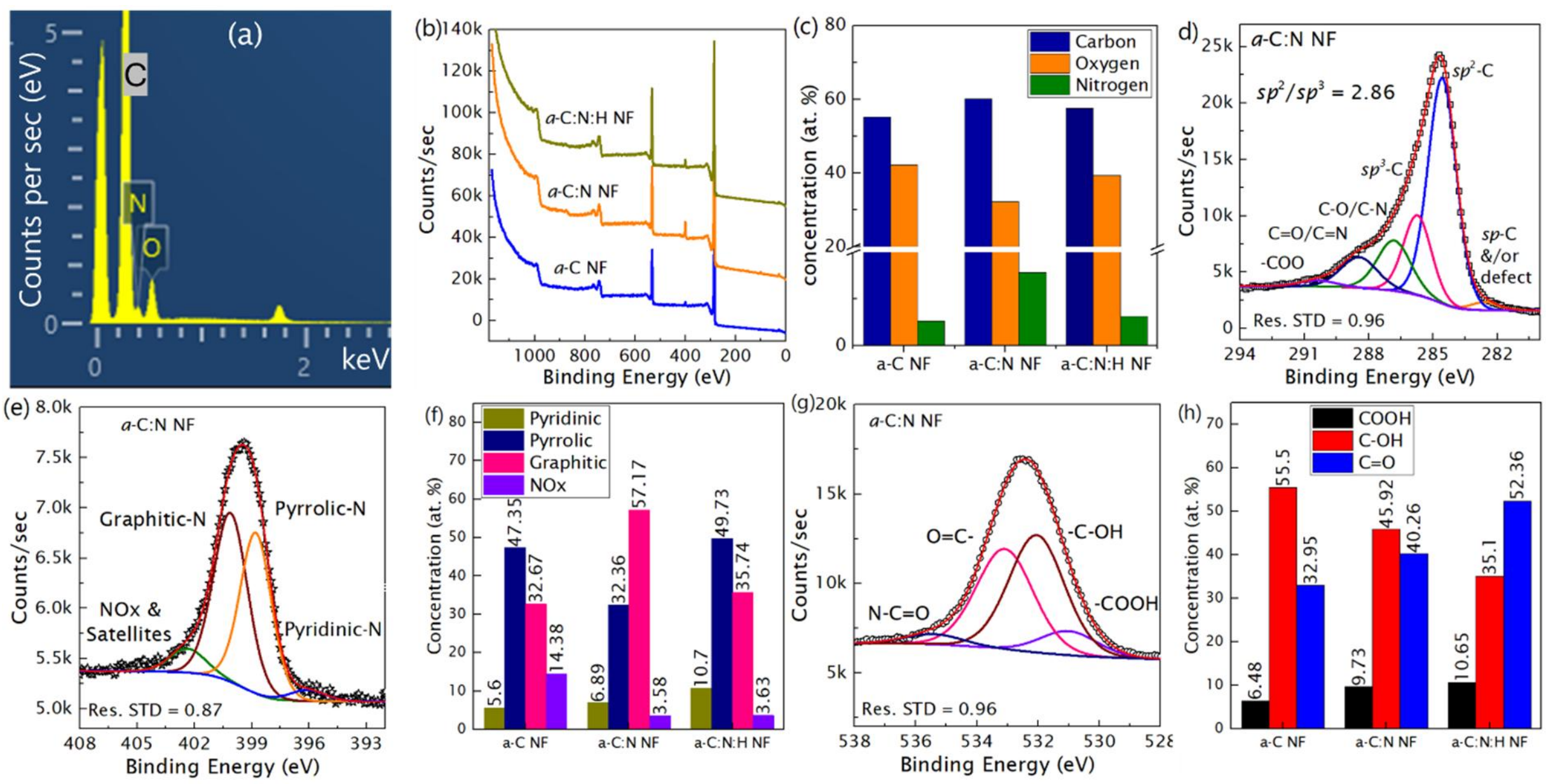

Figure 4: (a) EDX spectra of *a*-C:N nanofoams. XPS (b) survey spectra, and (c) plot of elemental compositions of all nanofoams. High resolution (d) C1*s*, (e) N1*s* spectra of *a*-C:N nanofoams with (f) relative concentration of N-bonding of all nanofoams. (g) O1*s* spectra of *a*-C:N nanofoams with (h) relative concentration of O-bonding of all nanofoams.

### 3.3. *Electrochemical investigation*

To exploit the NFs for potential applications, charge-storage performance was investigated and summarized in Table 1. Prior to the measurement, the voltage of device was optimized based on the cyclic voltammograms at different voltages keeping the scan rate fixed (Figure S5 of supplementary file). Figure 5a shows the cyclic voltammograms of *a*-C:N NF at different scan rates of 20 to 1000 mV/s, whereas cyclic voltammogram of *a*-C and *a*-C:N:H NF are supplied in Figure S6(a-b) supplementary file. A quasi-rectangular cyclic voltammogram and unchanged shape at higher scan rate indicates near-ideal supercapacitor behavior of NFs. Since the average thickness of the NFs prepared under different background gases (Figure 1d-f) are different, the volumetric cyclic voltammogram of all NFs is compared at 500 mV/s and shown in Figure 5b. However, the shape of cyclic voltammogram of *a*-C:N is comparatively found to be better in terms of quasi-rectangularity than that of *a*-C NF (Figure 5b). Both the areal capacitance and volumetric capacitance of the NFs are estimated from cyclic voltammogram at different scan rates and plotted in Figure 5c. The estimated areal capacitance of *a*-C, *a*-C:N, and *a*-C:N:H NF devices is 2.7, 4, and 1.2 mF/cm$^2$ at 20 mV/s, respectively. The areal capacitance of carbon NFs-based symmetric supercapacitor device obtained from our study is found to be much higher than the device made of pristine (0.16 mF/cm$^2$ at 100 mV/s) and functionalized vertical graphenes (1.6 mF/cm$^2$)[41], *sp*-carbon atomic wires wrapped polymer (2.4 mF/cm$^2$ at 20 mV/s),[42] CdS coated ZnO nanorods (0.03 mF/cm$^2$ at 50 mV/s),[43] $TiO_2$ nanogrid (0.74 mF/cm$^2$)[44] and titanate hydrate nanogrid (0.08 mF/cm$^2$).[44], etc.

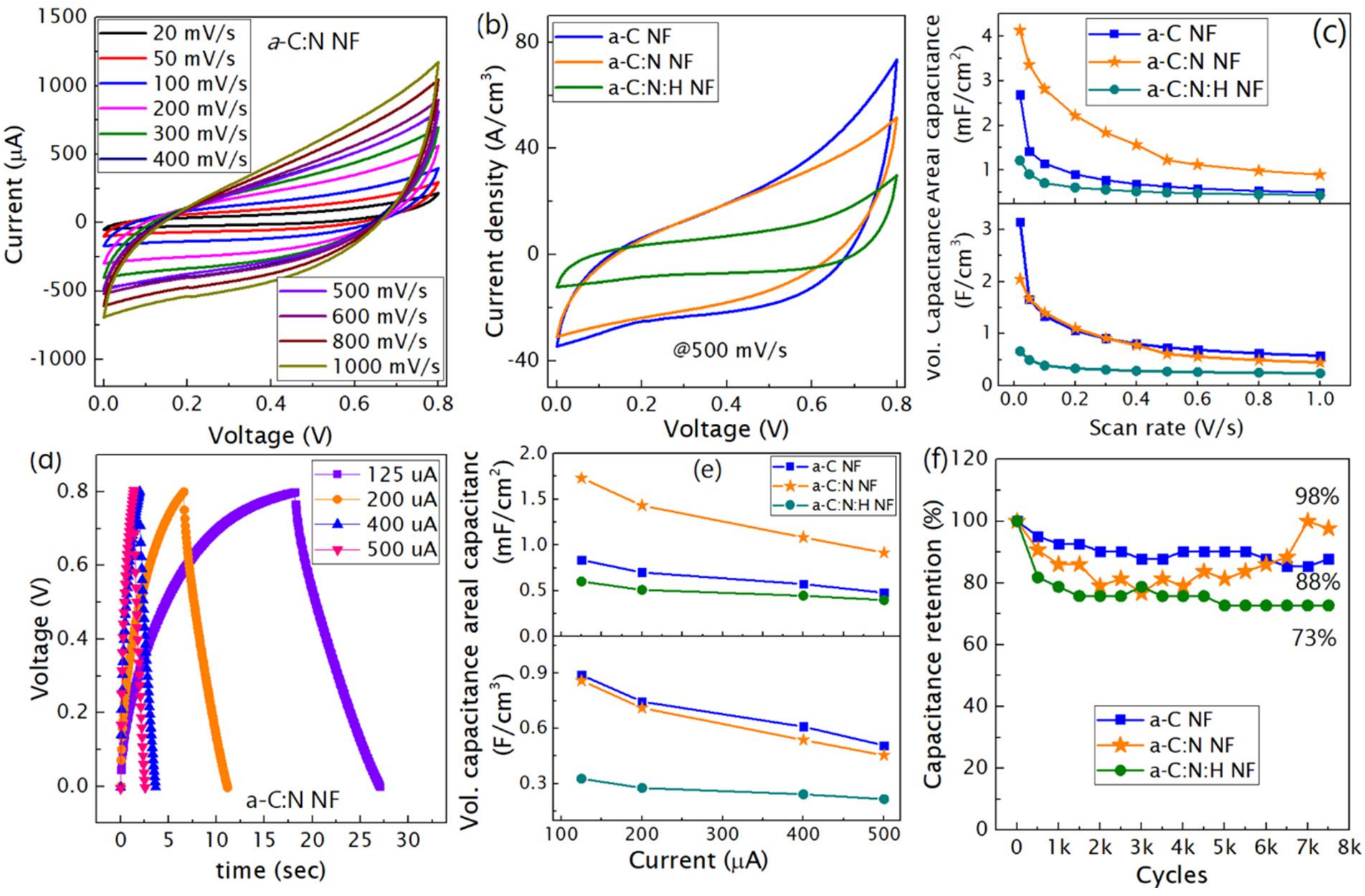

Figure 5: (a) Cyclic voltammogram of *a*-C:N nanofoams at different scan rates of 20 to 1000 mV/s. (b) Comparative volumetric cyclic voltammogram and (c) areal and volumetric capacitance of *a*-C, *a*-C:N and *a*-C:N:H nanofoams. (d) Charge-discharge profile of *a*-C:N nanofoams at different currents. (e) Plot of volumetric and areal capacitance of nanofoams at different currents. (f) Cycle stability of nanofoams.

While compared volumetrically, *a*-C NF delivered relatively higher specific capacitance (3.15 F/cm$^3$ at 20 mV/s) than the *a*-C:N NF (2.05 F/cm$^3$) but the rate capacitance at 1000 mV/s for *a*-C and *a*-C:N NF is found to be 18% and 22%, respectively. The NF-based aqueous symmetric device exhibits comparable volumetric capacitance of supercapacitor device reported in the literature such as FEG/$MoO_{3-x}$//FEG (3.96 F/cm$^{-3}$ at 1 mA/cm$^{-2}$, where FEG stands for functionalized partially exfoliated graphite),[45] $Fe_2O_3$ nanorod//$MnO_2$ (1.2 F/cm$^3$ at 0.5 mA/cm$^2$),[46] hydrogen-treated titania (H-$TiO_2$)@carbon nanowires//H-$TiO_2$@$MnO_2$ nanowires (0.71 F/cm$^3$ at 10 mV/s),[47] and reduced graphene oxide//$MnO_2$ nanorod (0.72 F/cm$^3$ at 10 mV/s).[48] The charge-discharge test of NFs at different current densities is also performed and shown in Figure 5(d) and Figure S6(c-d) in supplementary file. The areal and volumetric capacitance are also estimated at different current values and plotted in Figure 5(e). The charge-discharge result of the NF-based symmetric supercapacitor devices agree with cyclic voltammogram result. The cycle stability test over 10000 charge-discharge cycles at constant current indicates excellent stability of *a*-C:N NFs compared to other studied NFs. Although *a*-C:N:H NF is thicker compared to *a*-C NF, the higher areal (and volumetric) capacitance of *a*-C:N NF is attributed to the higher $sp^2/sp^3$-content, the presence of lower amount of COOH group in the structure, the higher N-content with higher graphitic-N compared and the relatively higher structural disorder, as discussed in XPS and Raman spectroscopic investigations (Table 1). In comparison to pristine *a*-C NF, better charge-storage performance of *a*-C:N NFs is due to the N-doping with graphitic-N and pyrrolic-N mostly. Incorporating N-doping enhances the density of states of $sp^2$-carbon at the Fermi level, and the associated capacitance so called quantum capacitance can be enhanced.[49] Among the N-configurations, positively charged graphitic-N attracts the electrolyte anions, contributes in double layer formation by structural disorder, and improve the charge-transfer kinetics.[28] Other N-functional groups, in particular pyrrolic-N and pyridinic-N, are well-known to contribute the pseudocapacitance in alkaline medium via the relations as below:

Pyrrolic-N (N-5)

$$\text{N-H} + OH^- \rightleftharpoons \text{N}^{\ominus} + H_2O + e^-$$

Pyridinic-N (N-6)

$$\text{N} + H_2O + e^- \rightleftharpoons \text{N-H}^+ + OH^-$$

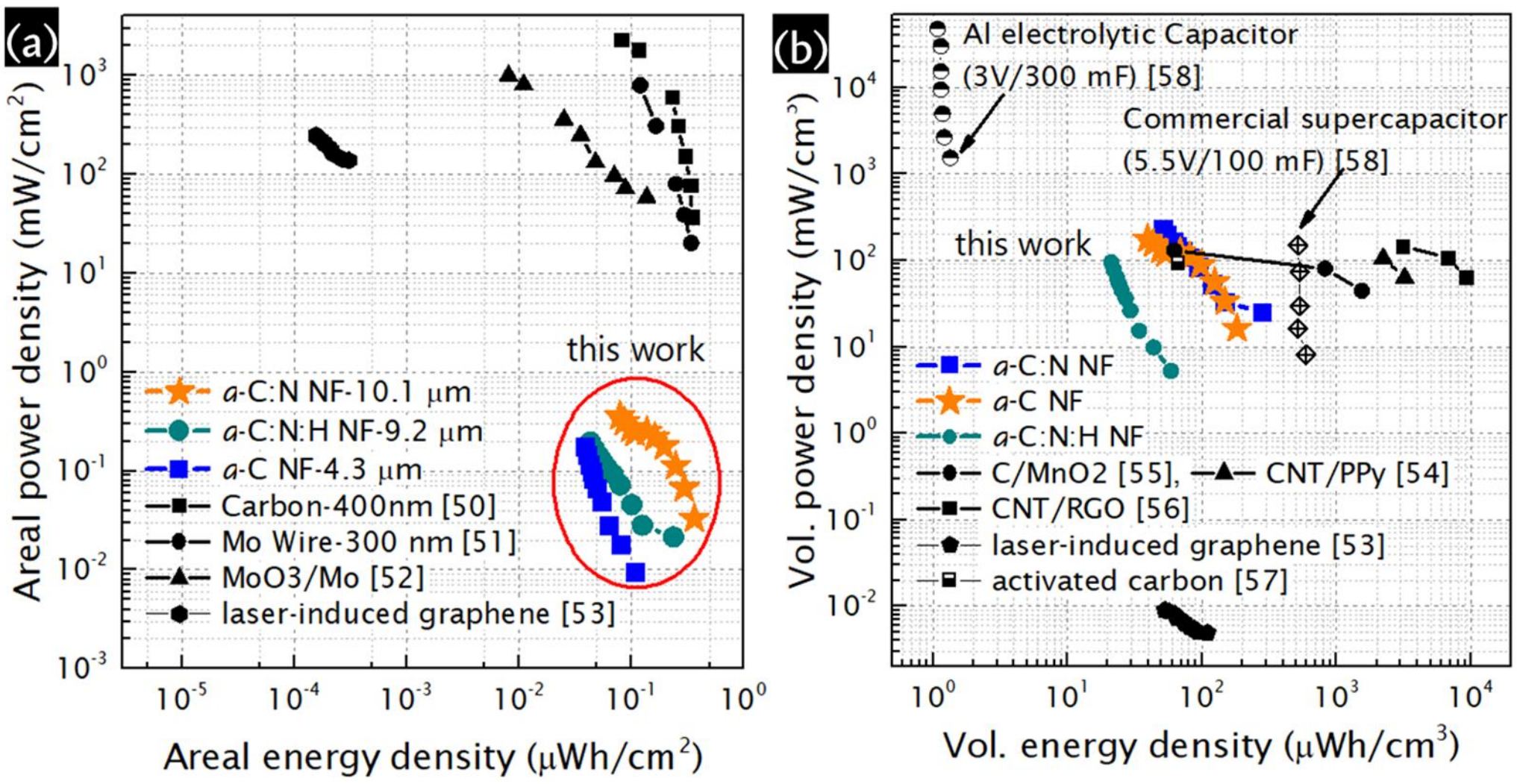

Figure 6: (a) Areal and (b) volumetric Ragone plot of carbon nanofoam supercapacitor devices comparing devices with others in literature. [CNT: carbon nanotube, PPy: Polypyrrole, RGO: reduced graphene oxide]

The areal and volumetric energy densities of symmetric supercapacitor at various areal and volumetric power densities are estimated and illustrated in Ragone plot (Figure 6a-b). The charge-storage capabilities of *a*-C NF supercapacitors are found to be comparable with the existing reports and commercial devices (Figure 6).[50][51][52][53][54][55][56][57][58] We would like to highlight that the thickness of the film in most of the reports is lower than that of a-C nanofoam. For example, thickness of carbon electrode is 400 nm [50] whereas the thickness for a-C NFs are in the range of 4-10 µm. Low thickness and low mass density often result in high charge-storage performance.[58] The performance of *a*-C NF supercapacitor device in terms of areal power density and volumetric energy density can be enhanced further by improving the NFs optimized in the present study, which will be communicated soon in our next article. In summary, the performance of *a*-C NFs ensures its potential application towards macro- to micro supercapacitor devices where the thickness of the electrode is limited in the range of 10 µm.[59]

### *3.4. Post-mortem analysis*

In order to check the morphology and structural quality of NFs, we have recorded the SEM images and Raman spectra of NFs before and after electrochemical investigations. For the after-electrochemical investigations, the NFs were washed with water several times to remove any residue of electrolyte and dried in an ambient environment. The morphology of *a*-C:N NFs before and after electrochemical investigations is shown in Figure 7(a) and 7(b), respectively. As can be seen, the NFs are agglomerated but still maintain the foam-like morphology. Similar observations have been seen for the other NFs studied here and hence not shown. Figure 7(c) is the linear background subtracted normalized Raman spectra of *a*-C:N NFs before and after the electrochemical test. Both spectra show the typical *a*-C features, and they almost overlapped on each other except for slight changes. To probe the slight changes, Raman spectra were deconvoluted as mentioned in the experimental sections and the result is shown in Table 2.

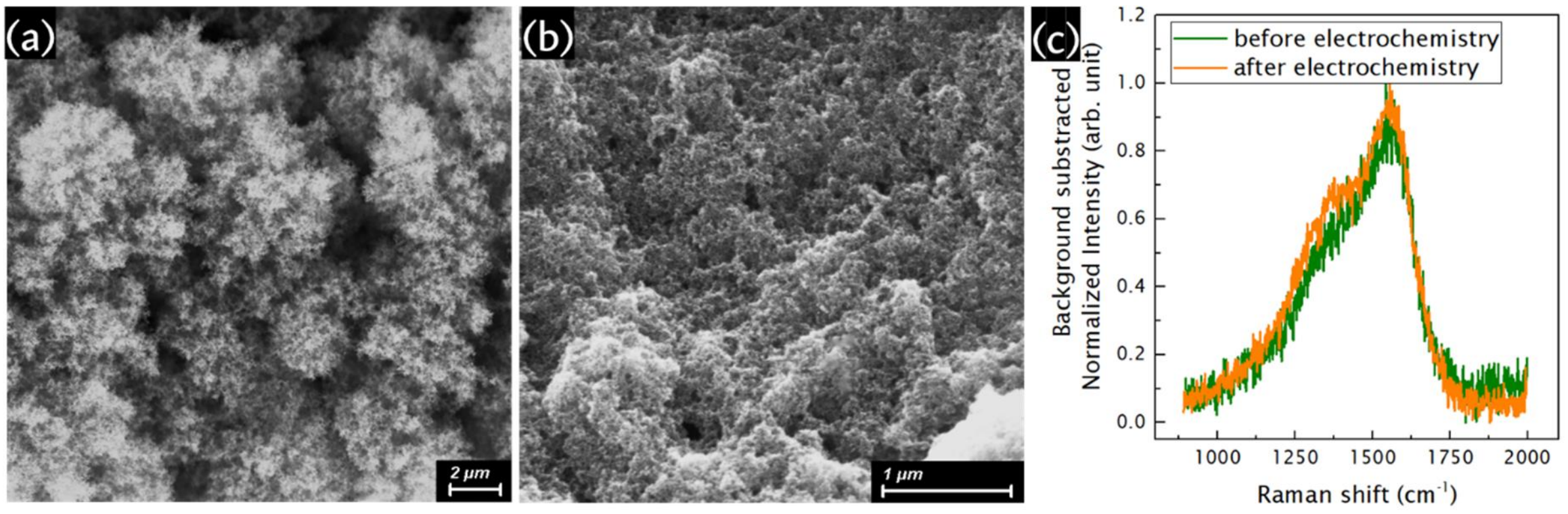

Figure 7: Post-mortem analysis. Scanning electron micrograph of *a*-C:N NFs (a) before and (b) after electrochemistry. (c) Raman spectra of *a*-C:N NFs before and after electrochemistry. Here the substrate is carbon paper.

Table 2: Raman extracted parameters of carbon NFs before and after electrochemical investigations

| | a-C:N NF | | a-C:N:H NF | | a-C NF | |
|---|---|---|---|---|---|---|
| | Before electrochemistry | After electrochemistry | Before electrochemistry | After electrochemistry | Before electrochemistry | After electrochemistry |
| G-peak position ($cm^{-1}$) | 1568 | 1562 | 1562 | 1573 | 1578 | 1565 |
| FWHM of G-peak ($cm^{-1}$) | 145 | 169 | 142 | 147 | 109 | 164 |
| $I_D/I_G$ | 0.78 | 0.65 | 0.4 | 0.54 | 0.65 | 0.57 |

The advantages of carbon NFs prepared by PLD are as follows: (i) the carbon-based active materials do not need any binder and conductive additives to fabricate the electrode; (ii) unlike the chemical vapor deposited graphene-based structure or carbon structure grown at high temperature (around 800 °C or more),[60] the deposition of carbon NF was conducted at room temperature and hence additional transfer process is not needed to use this materials for flexible supercapacitor applications; (iii) unlike the other carbon nanostructures, post treatment to improve the wettability (hydrophobic to hydrophilic) and hence charge-storage performance is not required[41] since the NFs are hydrophilic [7]; and (iv) N-incorporation using pulsed laser deposition is carried out at room temperature without a post-treatment in the presence of harmful $NH_3$, thanks to the out-of-equilibrium growth conditions achieved during PLD. N-incorporation in the carbon matrix is usually challenging unless high temperature growth of nanocarbon with a N-precursor[5][14] or post-deposition annealing treatment under caustic and unsafe $NH_3$ at high temperature is performed. Table 3 highlights the *a*-C NFs electrode preparation and compares the properties with other similar carbon nanostructures.

Table 3: Comparison of preparation method and properties of similar carbon nanostructures

| Carbon structures | Electrode fabrication strategy | Properties |
| --- | --- | --- |
| Carbon/Graphene nanosheets/nanowalls [33] | Plasma enhanced chemical vapor deposition. Can be grown directly on any substrate at the temperature of 750 °C | mass density = 535 mg/cm$^3$, Porosity – 71%, hydrophobic |
| Ultrathin carbon NF [8] | Naphtalene-Mediated Hydrothermal Sucrose Carbonization. growth temperature = 155 °C for 5 hrs. Need to wash the product and further processing is needed to use as electrode | mass density = 85 mg/cm$^3$ |
| Carbon aerogels [61] | Pyrolysis of resorcinol–formaldehyde aerogel at temperature of 600–1100 °C | surface area (400–800 m$^2$/g) and ultrafine cell/pore size (<100 nm), bulk densities of 0.55-0.60 g/cm$^3$ |
| Graphene aerogel [62] | Freeze casting at -100 °C followed by freeze-drying and thermal reduction at 800 °C under Ar environment | Lightweight, high porosity and high surface area |
| CNT scaffolds [63] | Pyrolysis at 1150 °C | Mass density = 20-240 mg/cm$^3$ |
| N,P-doped *a*-C foam (char) [5] | combustion of epoxy-based intumescent composite followed by annealing at 600–900 ºC in argon.<br>The char was dispersed in 1 mL of isopropanol, and 50 µL of Nafion for ink preparation | Density = 1.7–2.0 mg/cc.<br>surface area of 970.6-2498 m$^2$/g. |
| Carbon NF[25] | Laser irradiation of GO target and transfer to desired substrate by water-transfer process. | - |
| N-doped porous carbon NF[13] | Freeze-drying the transparent gel under liquid nitrogen and heat treatment at 600 ºC followed by $NH_3$ treatment at 400 ºC. Standard way the slurry was prepared to fabricate the electrode | Porous structure with measured surface area around 350 m$^2$/g |
| Pristine and N-containing *a*-C NF [This work] | Pulsed laser deposition,<br>It can be grown directly on desired substrate at room temperature. | mass density = 30-50 mg/cm$^3$, Porosity > 90%. hydrophilic |

## 4. Conclusion

In summary, a nitrogenated amorphous carbon nanofoam (*a*-C:N NF) directly on the current collector is synthesized by pulsed laser deposition at room temperature using nitrogen as a background gas. The *a*-C NF is highly porous with 98% volumetric void fraction, very lightweight with mass density of around 43 mg/cm$^3$, and low range order. To validate the influence of nitrogen in *a*-C matrix, the deposition of NFs was also carried out using Ar and $N_2$-$H_2$ background gases, while keeping other deposition parameters constant. Without the use of any binder and conductive-additives, the as-grown NFs are employed as supercapacitor electrodes. It has been seen that *a*-C:N NF with higher average thickness and more yield, higher $sp^2/sp^3$ content, more disordered structure, and higher N-content with more graphitic-N delivered better charge-storage performance compared to the other studied NFs. The aqueous symmetric supercapacitor device exhibited a high areal capacitance of 4.1 mF/cm$^2$ at 20 mV/s in aqueous electrolyte, excellent retention at high current, and 98% stability over 10000 charge-discharge cycles.

## Supporting information.

Additional SEM, XPS, Raman spectra, fitted curves with fitting details, and electrochemical results are provided in supporting information.

## Authors contributions

S.G. and C.S.C. planned and conceptualized the work. S. G. and A.M. performed the PLD deposition. S. G. did the Raman spectroscopy of samples and interpreted the result with V.R. S. G. and M.R. did the electrochemical measurements and assisted in the analysis. F.G. and G. B. did the XPS measurements. S.G. wrote the draft and all co-authors revised the manuscript and approved the final version of the manuscript.

## Notes

The authors declare no competing financial interest.

## Data Availability statement

All the data of this study are available in the main manuscript and the Supplementary Information.

## ACKNOWLEDGEMENT

S.G thank Horizon Europe (HORIZON) for the Marie Sklodowska-Curie Fellowship (grant no. 101067998-ENHANCER). Carlo S. Casari acknowledges partial funding from the European Research Council (ERC) under the European Union's Horizon 2020 Research and Innovation Program ERC Consolidator Grant (ERC CoG2016 EspLORE Grant Agreement 724610, website: www.esplore.polimi.it). Carlo S. Casari also acknowledges funding by the project funded under the National Recovery and Resilience Plan (NRRP), Mission 4 Component 2 Investment 1.3 Call for Tender 1561 of 11.10.2022 of Ministero dell'Università e della Ricerca (MUR), funded by the European Union NextGenerationEU Award Project Code PE0000021, Concession Decree 1561 of 11.10.2022 adopted by Ministero dell'Università e della Ricerca (MUR), CUP D43C22003090001, Project "Network 4 Energy Sustainable Transition (NEST)". A. M. acknowledges the *Energy for Motion* project of the Department of Energy of Politecnico di Milano, funded by the Italian Ministry of Education, University, and Research (MIUR) through the *Department of Excellence* grant 2018-2022.

# Supplementary Material

## Low-density functionalized amorphous carbon nanofoam as binder-free Thin-film Supercapacitor electrode

Subrata Ghosh[1,2*], Massimiliano Righi[1], Andrea Macreili[1], Francesco Goto[3], Marco Agozzino[1], Gianlorenzo Bussetti[3], Valeria Russo[1], Andrea Li Bassi[1], Carlo S. Casari[1*]

[1] *Micro and Nanostructured Materials Laboratory – NanoLab, Department of Energy, Politecnico di Milano, via Ponzio 34/3, Milano, 20133, Italy*

[2] Warsaw University of Technology, Faculty of Mechatronics, św. A. Boboli 8, Warsaw 02-525, Poland

[3] *Solid Liquid Interface Nano-Microscopy and Spectroscopy (SoLINano-Σ) lab, Department of Physics, Politecnico di Milano, Piazza Leonardo da Vinci 32, 20133 Milano, Italy*

Corresponding author email: subrata.ghosh@polimi.it (S.G.) and carlo.casari@polimi.it (C.S.C.)

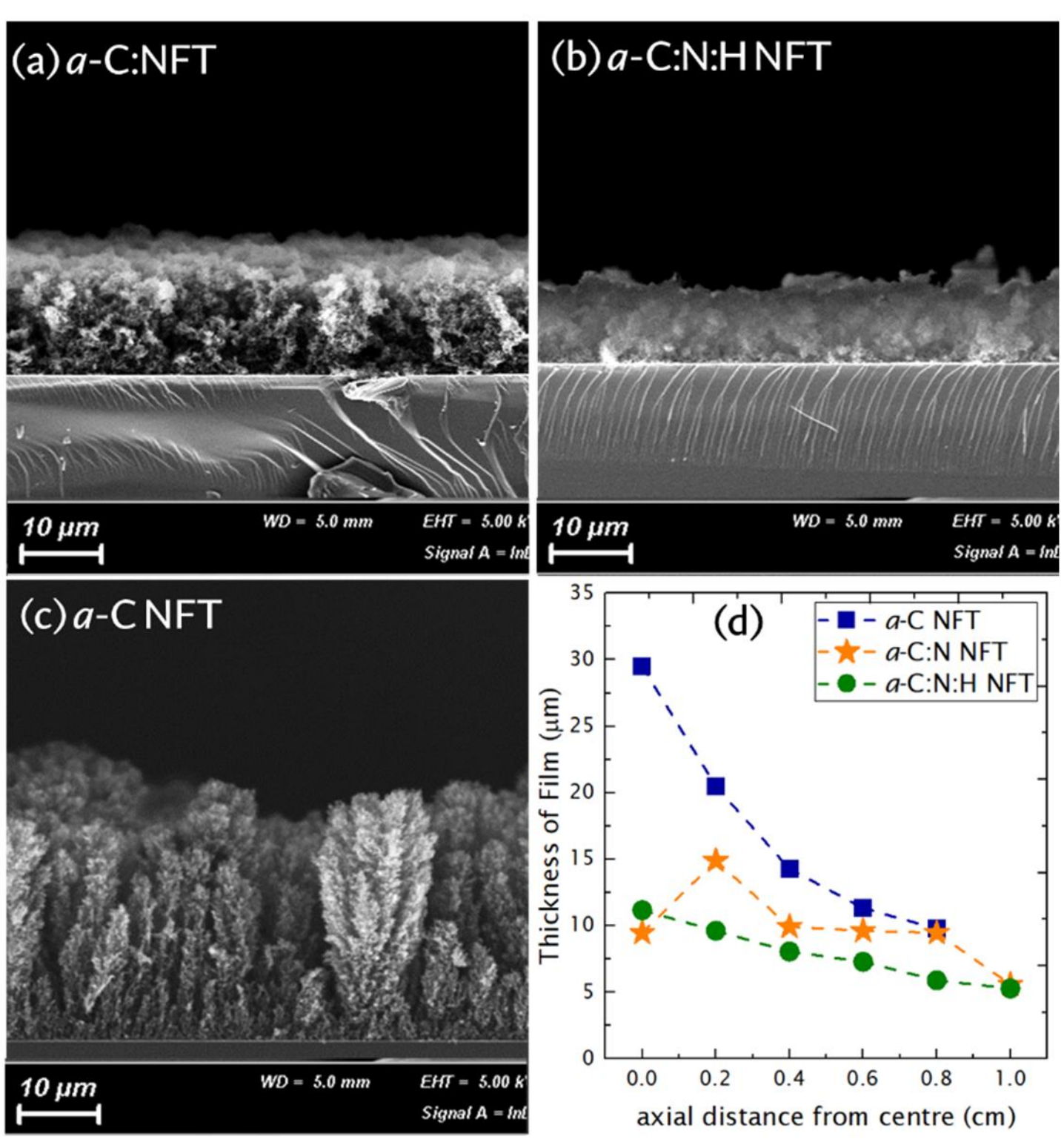

Figure S1: Cross-sectional micrograph of (a) *a*-C:N NFT, (b) *a*-C:N:H NFT, and (c) *a*-C NFT. (d) Plot of thickness of those nanofoams versus axial distance from centre. NFT: stands for nanofoam deposited in target-side.

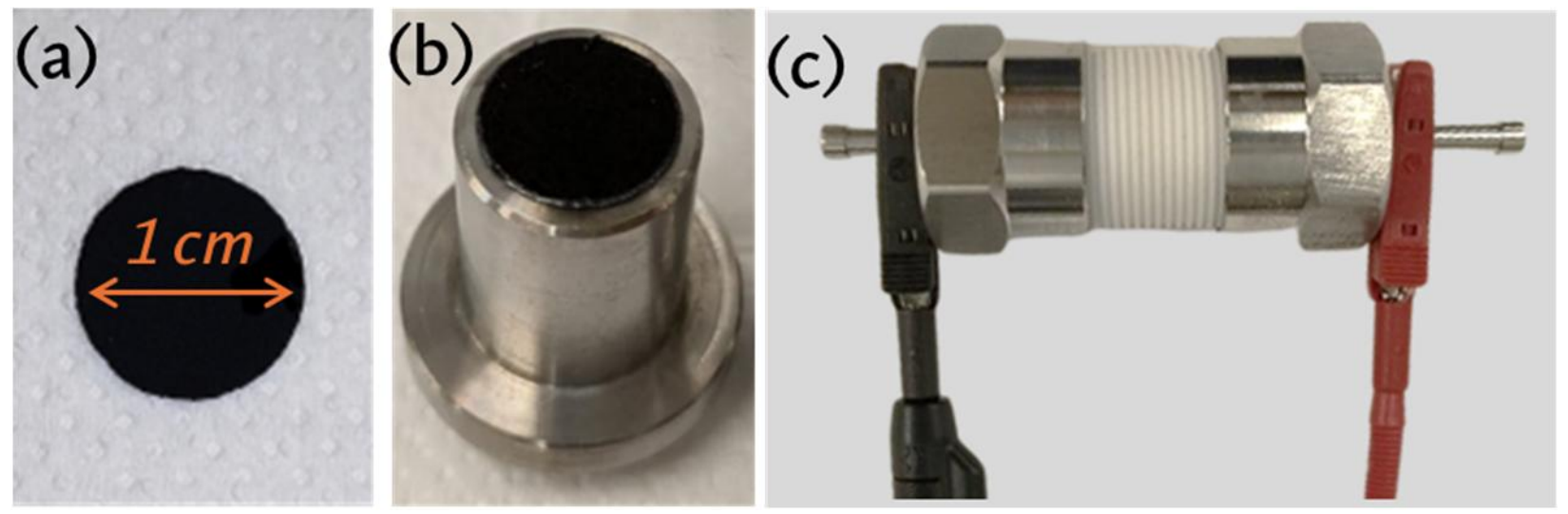

Figure S2: Photograph of (a) carbon nanofoam grown on carbon paper, (b) nanofoam-coated carbon paper on Swagelok cell die, and (c) assembled Swagelok symmetric supercapacitor device.

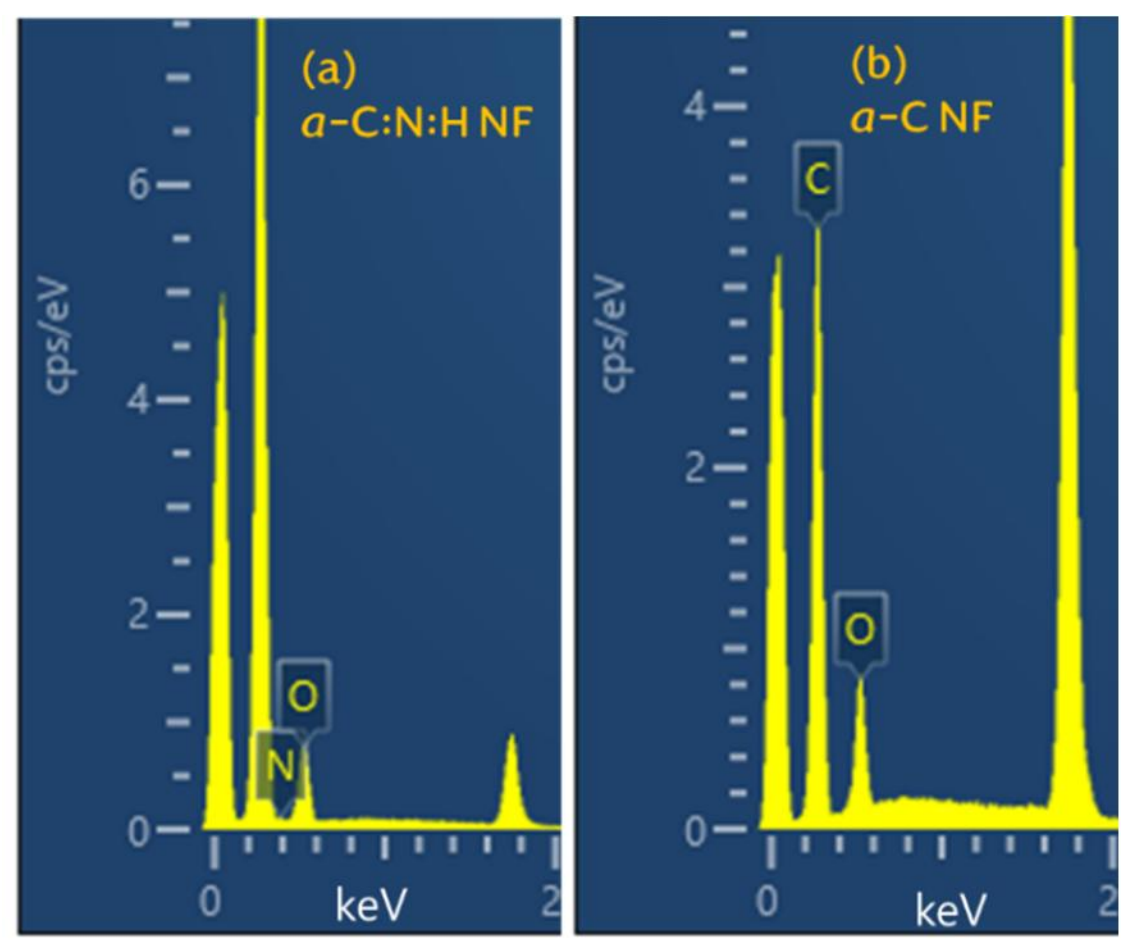

Figure S3: EDX spectra of (a) *a*-C:N:H and (b) *a*-C nanofoams

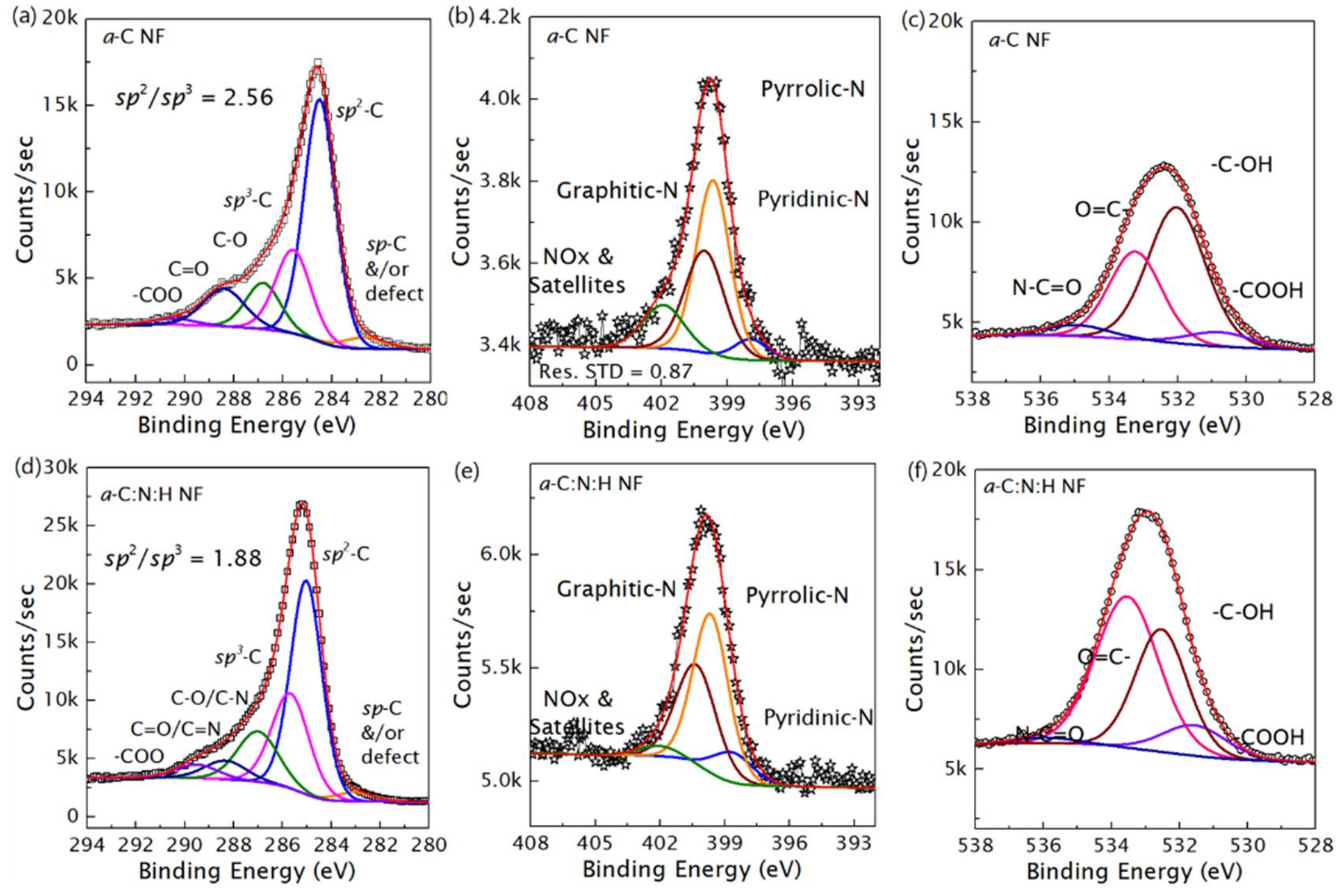

Figure S4: High resolution (a) C1*s*, (b) N1*s* and (c) O1*s* spectra with deconvoluted peaks of *a*-C nanofoams. High resolution (d) C1*s*, (e) N1*s* and (f) O1*s* spectra with deconvoluted peaks of *a*-C:N:H nanofoams.

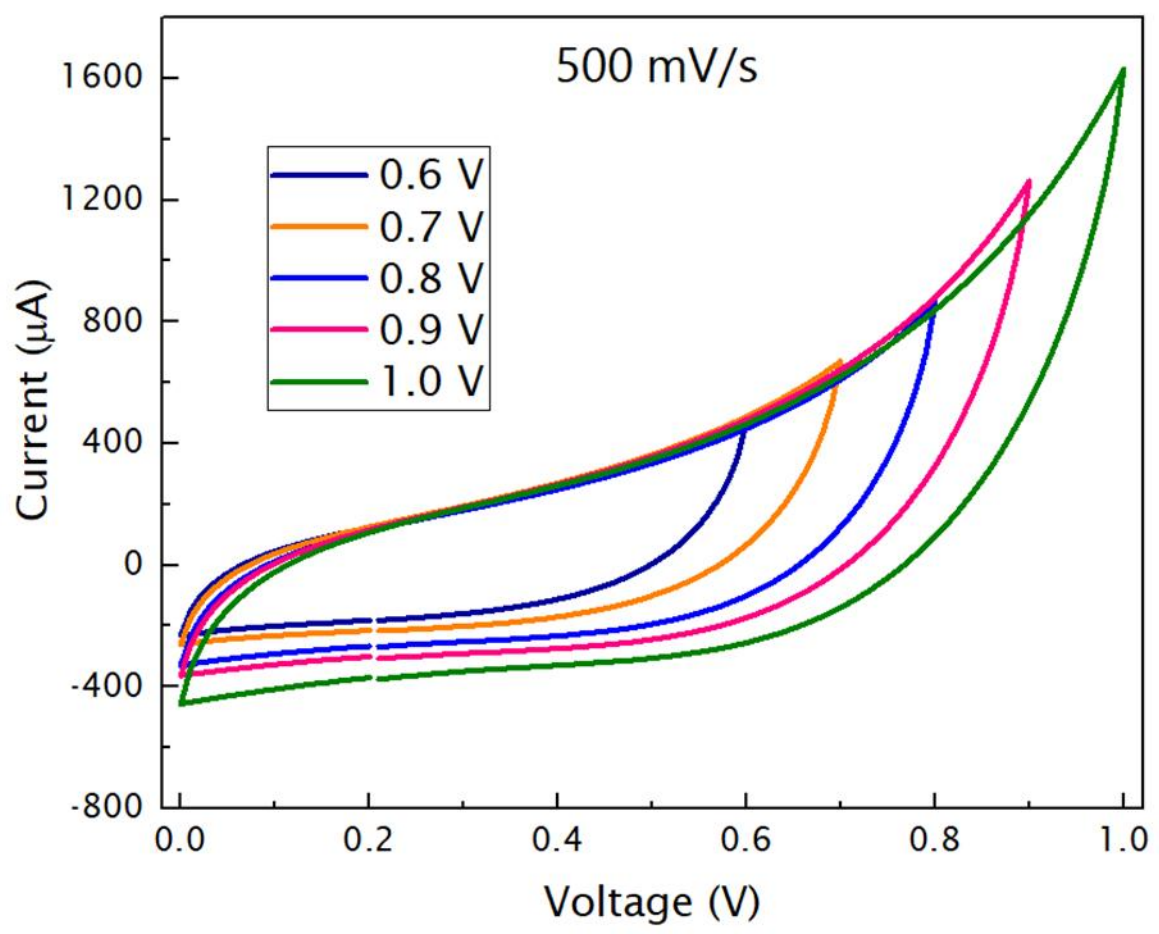

Figure S5: Cyclic voltammogram of *a*-C:N at different voltages.

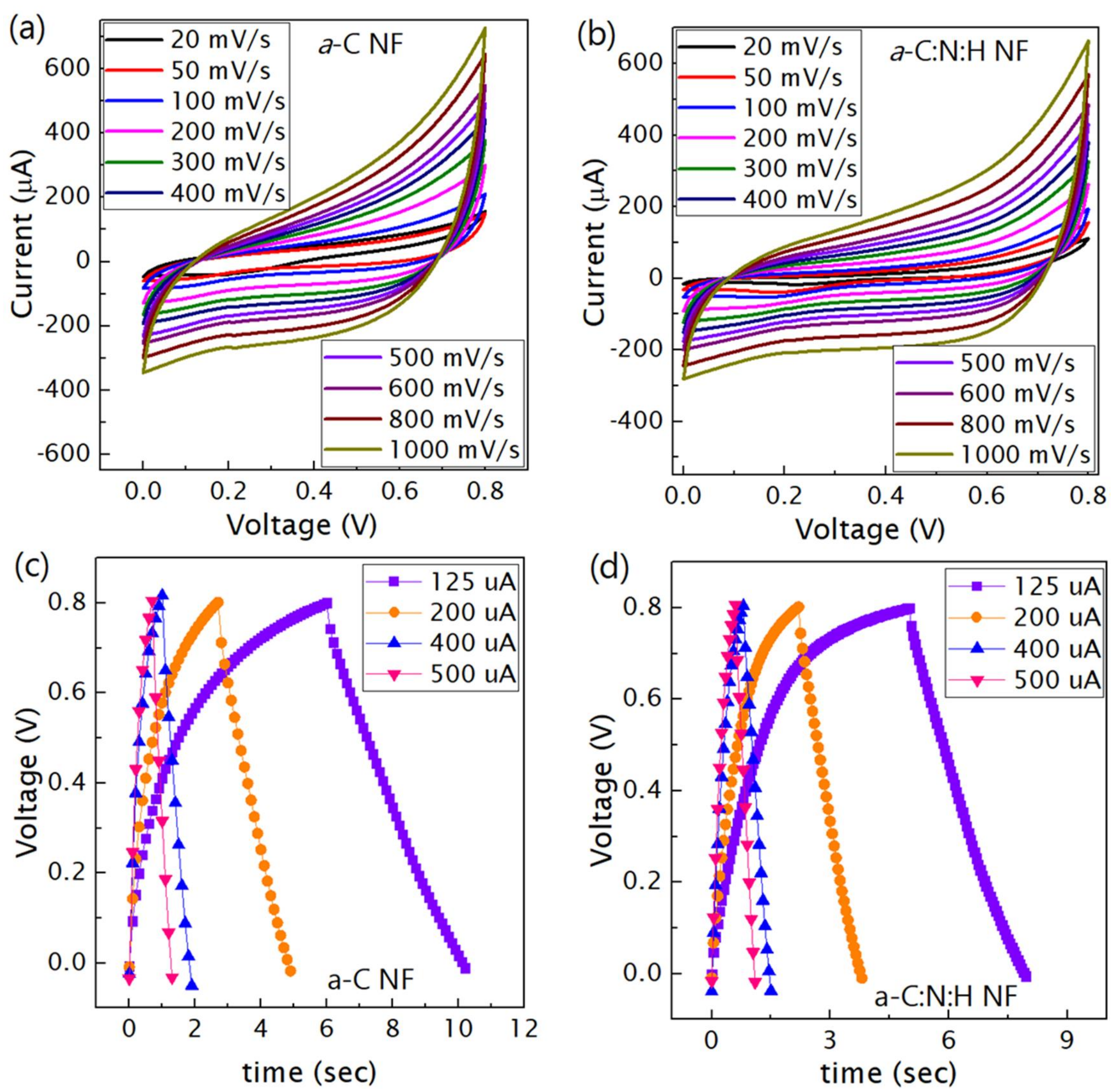

Figure S6: Cyclic voltammogram of (a) *a*-C and (b) *a*-C:N:H nanofoams at different scan rates of 20 to 1000 mV/s. Charge-discharge profile of (c) *a*-C, and (d) *a*-C:N:H nanofoams at different currents.